\documentclass[twocolumn,aps,prl]{revtex4}
\usepackage{amsmath}
\usepackage{graphicx}

\begin{document}

\title{Intersubband Electron Interaction in 1D-2D Junctions}

\author{V.A.~Sablikov}

\affiliation{Kotel'nikov Institute of Radio Engineering and Electronics,
Russian Academy of Sciences, Fryazino, Moscow District, 141190,
Russia}

\begin{abstract}
We have shown that the electron transport through junctions of one-dimensional and two-dimensional systems, as well as through quantum point contacts, is considerably affected by the interaction of electrons of different subbands. The interaction mechanism is caused by Friedel oscillations, which are produced by electrons of the closed subbands even in smooth junctions. Because of the interaction with these oscillations, electrons of the open subbands experience a backscattering. The electron reflection coefficient, which describes the backscattering, has a sharp peak at the energy equal to the Fermi energy and may be as high as about 0.1. This result allows one to explain a number of available experimental facts.
\end{abstract}

\maketitle

Quantum wires and quantum point contacts are of prime interest as model systems for studying the effects of electron-electron interaction, which is known to play an important role in one-dimensional systems, giving rise to a correlated state. Strong evidence in favor of the Luttinger liquid was obtained from the studies of tunnel-coupled quantum wires~\cite{Auslaender,Tserkovnyak}. However, experiments with isolated quantum wires do not directly testify to the presence of a Luttinger liquid but reveal many transport features that have not yet found any adequate explanation. Presumably, these features are related not only to electron-electron interaction in a wire, but also to the fact that the wire is connected with electron reservoirs. The most prominent and most discussed feature is the anomalous conductance plateau at $0.7 \times 2e^2/h$\cite{Thomas}. Other experimental facts can be classed into three groups:

(i) Experiments testifying to the electron localization above the potential barrier formed by a smooth (on the Fermi wavelength scale) electrode potential~\cite{Cronenwett, Auslaender2}. Such a localization is supposed to interpret the $0.7 \times 2e^2/h$ anomaly in terms of the Kondo effect~\cite{Cronenwett, Meir}. The localization mechanism remains unknown. Numerical calculations~\cite{Hirose} performed to justify the spin localization assumption have little force, because they use geometric dimensions of the contact that are close to the Fermi wavelength or even smaller.

(ii) Studies of the nonlinear conductance at a small (compared to the intersubband and Fermi energies) voltage applied along the quantum wire. An increase in the height of the differential conductance plateau is observed, whereas the ballistic conductance theory~\cite{Kouwenhoven, Martin} predicts its decrease with voltage and the inclusion of the electron-electron interaction via the selfconsistent ﬁfield does not qualitatively change this conclusion\cite{Shchamkhalova}. Moreover, in the experiment, the ﬁfirst conductance plateau rises to a level even higher than  $2e^2/h$~\cite{dePicciotto, Hansen}, which points to the appearance of an additional transport channel, e.g., through higher size-quantization subbands.

(iii) Observation of a specificﬁc scattering in the regions of sufficiently smooth transitions between one- and two-dimensional electron systems (1D-2D junctions). The scattering manifests itself as an effective resistance (estimated as $\sim 0.1h/2e^2$) connected in series with the quantum point contact~\cite{Hansen} or as the effect of the potential profile of the junction on the structure of the $0.7 \cdot 2e^2/h$ anomaly~\cite{Reilly}.

The present paper shows that these experiments can be explained (at least qualitatively) if one takes into account the interaction between electrons of different subbands in the junction between the 1D and 2D parts of the structure and, primarily, the interaction of the electrons passing through the junction with the electrons of the closed subbands. The interaction mechanism is related to the Friedel oscillations of electron density, which occur in the junction because of the reflection of higher subband electrons not passing through the contact. The physical picture is as follows: numerous electrons in the reservoirs collide with the contact, but only a small number of them (electrons belonging to the open subbands) can pass through the contact. All the other electrons are backscattered, causing Friedel oscillations of electron density. These oscillations evidently have different phases in different subbands. The phases depend on the form of the junction, but, since this form is described by a regular function, the summation over the subbands does not lead to the disappearance of oscillations. Away from the contact, the oscillations have a clearly pronounced component with a wave vector $2k_F$ ($k_F$ is the Fermi wave vector in the 2D reservoir). This assumption is supported by the experiment, which reveals the oscillatory structure of the electron density distribution at a large distance from the contact with the use of a probe microscopy technique~\cite{Topinka}. Our calculations show that the interaction of the electrons passing through the contact with the Friedel oscillations leads to a fairly strong backscattering. With this fact taken into account, it is possible basically to explain the experiments mentioned above.

\begin{figure}
\centerline{\includegraphics[width=8cm]{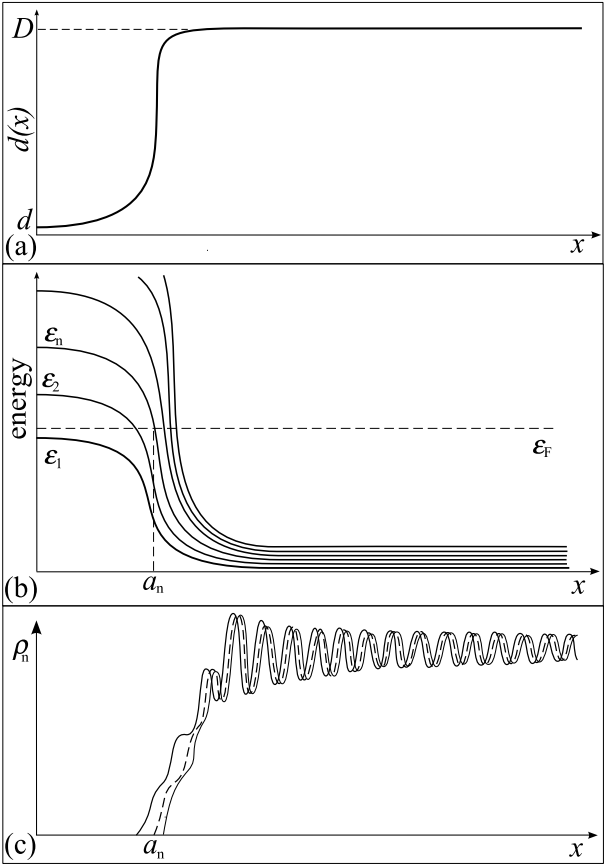}}
\caption{Electron density oscillations in a smooth 1D-2D junction: (a) the strip width variation in the junction, (b) the
size quantization subbands, and (c) the electron density oscillations in the closed subbands ($n=2,3,4\dots$).}
\end{figure}

Let us consider a 1D-2D junction in the form of a strip (Fig.~1) whose width $d(x)$ monotonically increases along the $x$ axis from $d \gtrsim k_ F^{-1}$ at $x = 0$ to $D\gg d$ at $x \to \infty$. The characteristic expansion radius $R$ considerably exceeds both $d$ and $k_F^{-1}$ . The problem consists in the evaluation of the reflection coefficient of electrons that are incident on the contact in the open subband and reflected as a result of their interaction with the Friedel oscillations caused by the electrons of the closed subbands. For calculations, we use the Born approximation, which is justified if the reflection coefficient is small. In the zero-order approximation with respect to the interaction, the wavefunctions can be determined in the framework of the standard adiabatic approximation (see, e.g., \cite{Glazman}). For the closed subbands,
we have
$$
\Psi_{n,k}(x,y)\simeq 2\chi_n(y) \sqrt{\frac{k}{k_n(x)} }\cos\left[\int^x_{a_n}\!\!dx' k_n(x') -\frac{\pi}{4}\right]\, ,
$$
where $n = 2, 3, \dots$ is the number of a subband; $k_n(x)$ is the wave vector of the longitudinal motion; $k =\lim_{x\to \infty} k_n(x)$; $\chi_n(y)$ is a transverse wavefunction; and $a_n(k)$ is the turning point. For simplicity, we assume that only one subband is open. For this subband the wave function is
\begin{equation}
\label{open_band}
\Psi_{1,k}(x,y)\simeq \chi_1(y) \sqrt{\frac{k}{k_1(x)}} \exp \left[i\int^x_{0}\!\!\!dx' k_n(x') \right]\, .
\end{equation}

To determine the potential of the perturbation causing the transition from the $|n,k\rangle $ state to the $|m,k'\rangle$ state, we use the Hartree–Fock approximation. Taking into account that the 1D-2D junction is smooth, we reduce the problem to effective one-dimensional equations by integrating the Hartree–Fock equations with respect to the transverse coordinates. As a result, we obtain the following expression for the reflection coefficient for the electrons in the open subband (i.e., for the $(1,k) \to (1,-k)$ transition):
$$
r_k=\frac{m}{i\hbar^2}\int \!dx\, \psi^*_k \hat{V} \psi_k\, ,
$$
where $\psi_k$ is the $x$-dependent part of the function $\Psi_{1,k}(x,y)$ in Eq.(\ref{open_band}). The perturbation potential contains the Hartree and exchange components: $\hat{V}=V^H+\hat{V}^{exc}$. The Hartree potential is the function
$$
V^H\!(x)\! =\! \sum_{n=2}^{N_f}\! \int \!\!dx' V^H_{1,n}(x,x') \rho_n(x') - \int\!\! dx' V^H_0(x,x') \rho_0(x')\, ,
$$
where $n(x)$ is the electron density in the $n$th subband and $\rho_0$ is the positive background charge density. The perturbation caused by the exchange interaction is described by the operator
$$
\hat{V}^{exc} \Psi (x) = - \sum_{n=2}^{N_f} \int dx' V^{exc}_{1,n}(x,x') \rho_n(x,x') \Psi(x') \, ,
$$
where $N_f$ is the index of the upper subband filled in the 2D part of the system and $\rho_n(x,x')$ is the density matrix, or, more precisely, its perturbation caused by the 1D-2D junction. The effective potentials $V^H_{n,m}(x, x')$ and $V^{exc}_{n,m}(x, x')$ for the direct and exchange interactions between the electrons belonging to the $n$th and $m$th subbands have the form
$$
V^H_{n,m}(x,x')=\int_0^{d(x)} \!\!dy\!\int_0^{d(x')}\!\! dy' V({\bf r},{\bf r'}) \chi_n^2(y)\chi_m^2(y') \, ,
$$
$$
V^H_0(x,x')=d(x')^{-1}\int_0^{d(x)}\!\! dy\!\int_0^{d(x')}\!\! dy' V({\bf r},{\bf r'}) \chi_1^2(y) \, ,
$$
\begin{eqnarray*}
V^{exc}_{n,m}(x,x')=\int_0^{d(x)}\!\!\! dy\!\int_0^{d(x')}\!\!\! dy' V({\bf r},{\bf r'})  \times \\
\chi_n(y) \chi_m(y) \chi_n(y') \chi_m(y')\, .
\end{eqnarray*}
Here, $V({\bf r}, {\bf r'})$ is the pair interaction potential, which is determined by the Coulomb interaction screened by metal electrodes, if any, and by two-dimensional electrons.

The calculation of the reflection coefficient can be simplified by taking into account the actual structure of the spatial distribution of electron density. Two density components are present: $\rho(x) \approx\bar{\rho}(x)+\tilde{\rho}(x)$, where the first component smoothly varies on the $k_F^{-1}$ scale and the second component oscillates with a wave vector of $\sim 2k_F$, with the oscillation amplitude smoothly (approximately as $x^{-3/2}$) decaying toward the depth of the 2D region. It is of special interest to consider the far zone ($x \gtrsim R$) where the wave vector of oscillations is close to $2k_F$, because these oscillations most efficiently scatter the electrons with Fermi energy in the backward direction. The contribution of the smooth component $\bar{\rho}(x)$ is small in terms of the parameter $(Rk_F)^{-1} \ll 1$. The contribution of the oscillating component $\tilde{\rho}(x)$ in the near zone is unimportant, because, here, the oscillation period is noticeably greater than the electron wavelength in the open subband. Thus, the problem can be simplified for the region $x \gtrsim R$ with allowance for the fact that the density oscillations of interest with the wave vector $ \sim 2k_F$ are produced by the electrons belonging to the lower subbands ($n\ll Dk_F/\pi$) 
and characterized by a longitudinal momentum close to $k_F$.

For such electrons, the interaction potentials  $V^H_{n,m}(x,x')$ and $V^{exc}_{n,m}(x,x')$ are simplified if the effective pair interaction radius $a$ is small compared to the wavelength in the transverse direction. In reality, this condition is satisfied in the far zone, because, in this zone, $a$ is on the order of the Bohr radius $a_B$ and the transverse wavelength is on the order of $D/n$. In this case, the difference between the potentials $V^H_{n,m}$ and $V^{exc}_{n,m}$ , as well as their dependence on the band indices, is insignificant. In addition, at distances greater than $a$, the interaction potential can be assumed to depend on the coordinate difference $|x-x'|$. Thus, we obtain
\begin{equation}
\label{potential}
V^H\approx V^{exc}\approx V(x-x') = \frac{2e^2}{\epsilon_0 D} U\left(\frac{|x-x'|}{a}\right)\, ,
\end{equation}
where $\epsilon_0$ is the permittivity of the semiconductor and $U(x)$ is the dimensionless potential, which depends on the screening in the system. In the case of the Coulomb interaction, the effect of screening by 2D electrons can be taken into account~\cite{Shekhtman} by using the permittivity of the 2D electron gas $\epsilon(q)$ in the random phase approximation~\cite{Stern}. In this case,
$$
U(x)=\frac{1}{2}\int_{-\infty}^{\infty}{\! \frac{dq}{|q| \epsilon(q)}e^{iqx}}\, .
$$

The oscillating electron density component $\tilde{\rho}(x)$ is determined by the summation over all of the closed subbands. If the electron-electron interaction is ignored, the oscillation amplitude caused by an individual subband decreases with distance as $1/x$. The inclusion of interaction leads to a slower decrease~\cite{Yue, Egger} of the amplitude. Because of the variance in the oscillation phases of different subbands, the total amplitude proves to decrease with distance faster than $1/x$. For example, in the case of a ﬂat edge of the 2D gas, the amplitude asymptotically decreases as $x^{-3/2}$\cite{Shekhtman}. The presence of the 1D-2D junction leads to a greater phase variance and, hence, to a decrease in the Friedel oscillation amplitude. In the region $x\gtrsim R$, for the electrons of the closed subbands, the phase at the Fermi level can be represented as
$$
\Phi (\gamma, x) \simeq k_Fx \sqrt{1-\gamma^2} - \Phi (\gamma,k_F) \, ,
$$
where $\gamma=n\pi/(k_FD)$. If $k_FD\gg 1$, one can assume that $\gamma$ is a continuous quantity and replace the summation over $n$ by integration with respect to $\gamma$. An important role is played by the second term $\Phi (\gamma,k_F)$ associated with the presence of the 1D-2D junction. In the absence of the junction, i.e., for a straight edge of the 2D region, $\Phi (\gamma,k_F)=0$. The specific form of the function $\Phi (\gamma,k_F)$ is determined by the form of the junction, but the general property of 1D-2D junctions is a sharp increase of the phase at $\gamma \to 0$, as it is shown in Fig.~2.

\begin{figure}
 \centerline{\includegraphics[width=8cm]{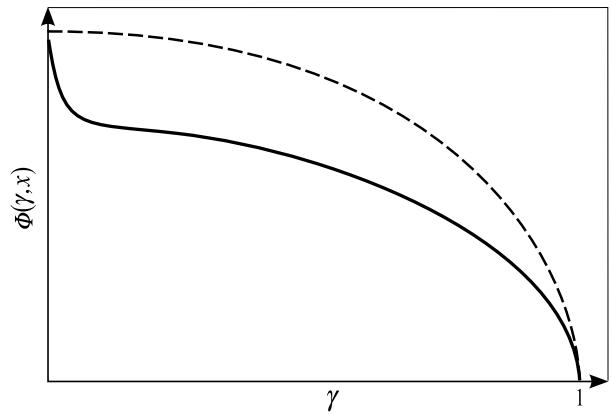}}
\caption{Phase vs. the band index (the solid curve) for the 1D-2D junction and (the dashed line) for the straight edge of the 2D electron gas.}
\end{figure}

To make the consideration more specific, let us consider the case specified by
\begin{equation}
\label{junc_form1}
d(x)=\left\{
\begin{array}{rl}
d / \sqrt{1-(x/R)^2}, & x<R \sqrt{1-\delta^2}\\
D, & x>R\sqrt{1-\delta^2}\,,
\end{array} \right.
\end{equation}
where $\delta=d/D \ll 1$. The characteristic feature of this $d(x)$ dependence is the presence of almost straight edges away from the transition region. Let us denote $\xi = x/R$. In the region $\xi>\sqrt{1-\delta^2}$,
$$
\Phi (\gamma, x) \simeq Rk_F\left[\xi - 1 + \delta^2/(3\gamma^2)\right]\,
$$

Calculating the density with the use of asymptotic expansions in $\lambda_F\equiv2Rk_F \gg 1$, we arrive at the following result:
\begin{eqnarray*}
\tilde{\rho}(x) \simeq -\frac{2k_F}{\pi}\frac{D}{4\pi R}\sqrt{\frac{\pi}{2\lambda_F}} \frac{\exp{-\lambda_F\delta\sqrt{2(\xi-1)/3}}}{\xi-1}\times \\
{\mathrm {Re}}\left[\frac{\exp{\left\{i[\lambda_F(\xi-1)-\pi/4]\right\}}}{\sqrt{\xi -1}+i\delta\sqrt{3/2}}\right].
\end{eqnarray*}
One can see that, unlike the case of the straight edge, in the 1D-2D junction, the density oscillations decay exponentially with a characteristic decay length of $\sim (D/k_Fd)^2/R$. For $D\gg R$, this length is much greater than $R$ and, hence, $k_F^{-1}$. Therefore, such a decay of oscillations is insignificant.

The reflection coefficient, correct to the phase factor, is expressed as
\begin{equation}
\label{reflection}
r_k\simeq \frac{i}{\pi \sqrt{\pi}}\frac{ak_F}{a_Bk}\left\{\!\left[\tilde{U}_{2ka}\!-\!\frac{\tilde{U}_0}{2}\right]\! F\!\left(\!\frac{k}{k_F\!}\right) - \Delta [z,U]\!\right\}\,,
\end{equation}
where the Fourier transform $\tilde{U}_q$ of the potential $U(t)$, which depends on the dimensionless coordinate $t=x/a$ is introduced; $\tilde{U}_{2ka}$ and $\tilde{U}_0$ are $\tilde{U}_q$ at $q=2ka$ and $q=0$. Note that, although the potentials $V^H$ and $V^{exc}$ (see Eq. (\ref{potential})) depend on $D$, the final expression for $r_k$ is free of this dependence because of the summation over the subbands. The function $F(k/k_F)$ has the form
$$
F(z)=\int_0^{\infty}\! dy \frac{\cos(2y\!+\!c\!-\!\pi/4)}{(y+c/2)^{3/2}}e^{2izy-2\delta\sqrt{\lambda_F (y+c/2)/3}} \,,
$$
where $c\sim 1$ is a constant arising because of the limitation imposed by the applicability of the asymptotic formulas for Friedel oscillations. In Eq.~(\ref{reflection}), the term $\tilde{U}_{2ka}$ is caused by the Hartree interaction, and the term $\tilde{U}_0$ is associated with the exchange interaction. The combination $[2\tilde{U}_{2ka}-\tilde{U}_{0}]$ appears in a standard way in the theory of electron scattering by Friedel oscillations~\cite{Rudin,Zala,Yue,Shekhtman}. The last term in Eq. (\ref{reflection}),
\begin{eqnarray*}
\Delta[z,U]=\!\int_0^{\infty}\! \! \! dy \frac{\cos(2y\!+\!c\!-\!\pi/4)}{(y+c/2)^{3/2}}e^{2izy-2\delta\sqrt{\lambda_F (y+c/2)/3}}\times \\
\int_{2y/ak_F}^{\infty}\!\!dt\, U(t),
\end{eqnarray*}
is also associated with the exchange interaction. It arises because of the difference in the direct and exchange intersubband interactions in the 1D-2D junction. Within the part of the junction, where the electron density of closed subbands is absent, the exchange interaction between the closed subband electrons and the electrons of the open subband is also absent, while the direct interaction extends over a distance of about $a$. For $ak_F<1$, the term $\Delta[z,U]$ vanishes. In any case, $\Delta[z,U]$ does not noticeably affect the reflection coefficient, so that $|r_k|$ can be estimated by the first term of Eq.~(\ref{reflection}).

The dependence of the reflection coefficient on the wave vector $k$ is mainly determined by the function $(k_F/k)F(k/k_F)$ plotted in Fig.~3. The characteristic feature of this function is the sharp peak at $k=k_F$, near which the function follows a root dependence on  $|k-k_F|$. The factor  $[\tilde{U}_{2ka}-\tilde{U}_0/2]$ varies more smoothly. If $U(t)=U_a \exp{(-|t|)}$, we have $\tilde{U}_0=2U_a$ and $\tilde{U}_{2ka}=2U_a/(1+4k^2a^2)$. In the case of a screened Coulomb interaction, $\tilde{U}_0\approx \pi/2$ and $\tilde{U}_{2ka}=\pi /[2(1+ka_B)]$. Hence, the dependence of $|r_k|$ on $k$ is approximately identical to the dependence shown in Fig.~3. The value of $|r_k|$ for actual values of the parameters $k_F$ and $a$ can be on the order of several tenths, and $|r_k|^2 \sim 0.1$. The contribution of the exchange interaction is predominant.

\begin{figure}
\centerline{\includegraphics[width=8cm]{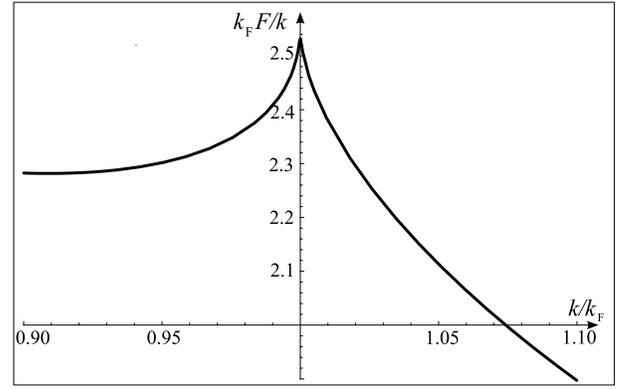}}
\caption{Function $(k_F/k)F(k/k_F)$ determining the dependence of the reflection coefficient on the electron energy; $\lambda_F = 30, \delta = 10^{-3}, c = 0.5$.}
\end{figure}

The results obtained above qualitatively hold for 1D-2D junctions of other forms. For example, in the case of a smoother junction with $d(x)=D-(D-d)\exp[-(x/R)^2]$, the phase $\Phi (\gamma,k_F)$ has a root singularity for small values of $\gamma$: $\Phi(\gamma,k_F) \simeq Rk_F(\xi-c_1\sqrt{\gamma})\sqrt{1-\gamma^2}$, where $c_1 \simeq 5/4$. In this case, a length parameter appears,  $l_c=4(c_1R)^4k_F^3$, which is considerably greater than $R$. Since the phase variance is stronger than that in the case considered above, i.e., in the case of $d(x)$ given by Eq.~(\ref{junc_form1}), the Friedel oscillation amplitude decreases but its dependence on distance weakens within the length $l_c$: $\tilde{\rho}\sim x^{-1}$.  For $x\gg l_c$, the dependence $\tilde{\rho}\sim x^{-3/2}$ holds. As a result, the electron reflection coefficient characterizing the scattering in the far zone decreases approximately by a factor of $\sim \pi^2(c_1\lambda_F)^{-2}\ln(c_1\lambda_F)$. However, at the same time, the size of the near zone increases and the role of the scattering processes in it  becomes more significant. This case requires special consideration.

Thus, the interaction between electrons of different subbands in 1D-2D junctions can be sufficiently strong to affect the electron transport in quantum point contacts and quantum wires. The effect of this interaction is as follows:

(i) The backscattering of electrons of open subbands leads to a decrease in conductance, which explains the presence of a specific resistance observed experimentally for 1D-2D junctions~\cite{Hansen}. The estimate obtained above, $|r_k|^2\sim 0.1$, agrees well with these experiments;

(ii) In a quantum wire connecting two electron reservoirs, the electron scattering occurs in two opposite junctions, which results in the appearance of quasibound states and enhances the effect of the scattering mechanism under discussion on the conductance;

(iii) Since the scattering by Friedel oscillations decreases with an increase in the electron energy for $k>k_F$, the differential conductance of the contact may increase with increasing bias voltage. In principle, the differential conductance may even exceed the value of $2e^2/h$, because the intersubband interaction creates a possibility for the transitions from the closed subbands to the open ones (even for smooth 1D-2D junctions). Indeed, the matrix element of the electron transition from the $n$th subband to the first one due to the interaction with the electrons of the $m$th subband is determined by the terms of the form $\langle\chi_1(y) \chi_m^2(y) \chi_n(y)\rangle\neq 0$. If, for example, $\chi_n(y)\sim \sin(\pi n y/D)$, such transitions are possible for all of the odd values of $n$.

This work was supported by the Russian Foundation for Basic Research (project no. 05-02-16854) and by the Russian Academy of Sciences (programs "Quantum Nanostructures" and "Strongly Correlated Electrons in Semiconductors, Metals, Superconductors, and Magnetic Materials").

\end{document}